\documentclass[12pt]{article}
\usepackage{graphicx}
\usepackage{cite}

\def \bea{\begin{eqnarray}}
\def \beq{\begin{equation}}
\def \b{{\cal B}}

\def \eea{\end{eqnarray}}
\def \eeq{\end{equation}}

\def \s{\sqrt{2}}

\def \3half{\frac{3}{2}}

\def\tl{\tilde\lambda}
\def\xNP{\xi^{\rm NP}}
\def\delNP{\delta^{\rm NP}}
\def\phiNP{\phi^{\rm NP}}
\def\xNPk{\xi^{\rm NP}_k}
\def\delNPk{\delta^{\rm NP}_k}
\def\ls{\lambda_{B_s}}

\def\l0{\lambda_{B^0}}

\begin{document}
\begin{flushright}
CERN-PH-TH/2007-169\\
September 2007 \\
\end{flushright}
\centerline{\bf  STUDYING NEW PHYSICS AMPLITUDES}
\medskip
\centerline{\bf IN CHARMLESS 
%$B$ and 
$B_s$ DECAYS}
\bigskip
\centerline{Robert Fleischer and Michael Gronau
\footnote{Permanent address: Physics Department, Technion-Israel Institute of Technology, 
Haifa, Israel.}}
\medskip
\vskip3mm
\centerline{\it Theory Division, Department of Physics, CERN}
\centerline{\it CH-1211 Geneva 23, Switzerland}
\bigskip
\centerline{\bf ABSTRACT}
\medskip
A method based on flavour SU(3) is proposed for identifying and extracting New 
Physics (NP) amplitudes in charmless $\Delta S=1$ $B_s$ decays 
using time-dependent CP asymmetries in these decays  and in flavour 
SU(3) related $\Delta S=0$ decays. For illustration, we assume a hierarchy, 
$\sim 1: \lambda : \lambda^2$ ($\lambda=0.2$), between a dominant $\Delta S=1$ 
penguin amplitude, a NP amplitude and a Standard Model amplitude with weak phase 
$\gamma$. An uncertainty from SU(3) breaking corrections, reduced by
using ratios of hadronic amplitudes,   is further suppressed by a factor $\lambda$. 
We discuss examples for pairs of decays into two neutral 
vector mesons, $B_s\to\phi\phi,~B_s\to\phi \bar K^{*0}$ and 
$B_s\to K^{*0}\bar K^{*0},~B^0\to K^{*0}\bar K^{*0}$, where the magnitude of the 
NP amplitude, its weak and strong phases can be determined.
\bigskip

\noindent
PACS Categories:  13.25.Hw, 11.30.Er, 12.15.Ji, 14.40.Nd

\bigskip
\centerline{\bf I.  INTRODUCTION}
\bigskip

Strageness-changing charmless $B$ and $B_s$ decays dominated by $b \to s$ 
penguin amplitudes, suppressed by CKM and loop factors, are sensitive to New 
Physics (NP) effects~\cite{Gronau:1996rv,Grossman:1996ke,Ciuchini:1997zp,
London:1997zk,Barbieri:1997kq}.
Virtual new heavy particles at a TeV mass scale may affect Standard Model 
predictions~\cite{London:1989ph,Gronau:1989ia,Grossman:1996ke,Fleischer:1996bv}, 
$C\equiv -A_{CP} \simeq 0 +{\cal O}(\lambda^2), S \simeq -\eta_{\rm CP}\sin 
2\beta + {\cal O}(\lambda^2)$, for  time-dependent asymmetries in decays to CP 
eigenstates with CP-eigenvalue $\eta_{\rm CP}$. Corrections of order $\lambda^2$, 
$\lambda\equiv |V_{us}| =0.2257$~\cite{Wolfenstein:1983yz,Yao:2006px}, are 
due to terms in decay amplitudes involving a weak phase  $\gamma$.
These SM predictions have been tested in a large class of processes including 
$B^0\to XK_S, X = \phi, \pi^0, \eta', \omega, \rho^0, f_0(980), K^+K^-, K_SK_S, \pi^0\pi^0$.
Asymmetries measured in the first two processes~\cite{HFAG}, 
 \beq
-\eta_{\rm CP}S(B^0\to \phi K_S) = 0.30\pm 0.17~,~~~~
-\eta_{\rm CP}S(B^0\to \pi^0K_S)= 0.38\pm 0.19~,
\eeq
indicate potential discrepancies (currently at levels of $2.2\sigma$ and $1.6\sigma$) 
with respects to the value $\sin 2\beta = 0.681\pm 0.025$ measured in 
$b\to c\bar cs$ transitions~\cite{HFAG}. 

Two techniques, based on QCD factorization and flavour SU(3), have been applied 
to control within the Standard Model corrections  of order $\lambda^2$ to $C$ and $S$.  
In an approach using QCD factorization~\cite{Buchalla:2005us,Beneke:2005pu,Cheng:2005bg} 
one calculates these terms from first principle at leading order in $1/m_b$ and $\alpha_s$. 
The calculations involve uncertainties partly due to penguin contractions, chirally-enhanced 
$1/m_b$ suppressed terms and nonperturbative input parameters. 
In a flavour SU(3) approach one relates these corrections to amplitudes for $\Delta S=0$ 
$B^0$ decays. Using measured rates for the latter processes one obtains upper bounds 
on these corrections~\cite{Grossman:2003qp,Gronau:2003ep,Gronau:2003kx}. 
The upper bounds involve uncertainties from SU(3) breaking corrections, usually assumed
to be of order $m_s/\Lambda_{\rm QCD}$.
Under conservative assumptions about strong phases, predictions of CP asymmetries 
in both methods involve theoretical uncertainties of order $\lambda^2$. 
This makes it extremely difficult to identify NP amplitudes if these are of order 
$\lambda^2$ relative to dominant penguin amplitudes. A question seeking an answer is 
how to identify and extract NP amplitudes, which in principle could be of order $\lambda$.
 
In the present Letter we propose a more precise, yet experimentally more
challenging method for controlling small Standard Model amplitudes based 
on flavour SU(3) symmetry. The method requires measuring time-dependent asymmetries 
in pairs of SU(3)-related $B^0$ and $B_s$ decays. Assuming given values for 
CKM phases, $\beta, \gamma$ and the phase $\phi_s$ of $B_s$--$\bar B_s$ mixing,
obtained for instance in  $B\to J/\psi K_s, B\to D^{(*)}K^{(*)}$ and $B_s\to J/\psi\phi$, 
respectively,
we suggest a way for analyzing and extracting NP decay amplitudes in a class of $b\to s$ 
penguin-dominated decays.  Our considerations do not depend on whether $\phi_s$ obtains
a NP contribution, thereby modifying the Standard Model prediction 
$\phi_s=-2\lambda^2\eta$~\cite{Wolfenstein:1983yz}.
This prediction can be tested in $B_s\to J/\psi\phi$~\cite{Dunietz:2000cr}.

A formulation similar to the one presented here, neglecting NP contributions, 
has been advocated in Ref.~\cite{Fleischer:1999pa} as a method for determining 
$\gamma$ in the U-spin pair of processes, $B_s(t)\to K^+K^-$ and $B^0(t)\to\pi^+\pi^-$.
Refs.~\cite{London:2004ws,Gronau:2007ut} contain earlier studies of NP effects 
in $b\to s$ penguin-dominated decays, involving different approaches and further 
assumptions about negligible strong phases of NP amplitudes. 

\bigskip
\centerline{\bf II. THE METHOD}
\bigskip

Consider a pair of $\Delta S=1$ and $\Delta S=0$ charmless decay processes for 
$B_s$ and $B^0$, respectively, where final states are related to each other by a 
U-spin transformation, $d\leftrightarrow s$.  Let us denote a potential NP amplitude 
in the first process by its magnitude $A_{\rm NP}$, CP-conserving phase $\delNP$ and 
CP-violationg phase $\phiNP$. Assuming no NP contribution in $\Delta S=0$ decays, 
the decay amplitudes for the two processes can be generally expressed as 
\bea\label{AmpBs}
A(B_s\to f)_{\Delta S=1} & = & A_1\left(1 + \xi' e^{i\delta'}e^{i\gamma} + 
\xNP e^{i\delNP}e^{i\phiNP}\right)~,\\
\label{AmpB0}
A(B^0\to Uf)_{\Delta S=0} & = & -\tl A_0\left(1 - \tl^{-2}\xi e^{i\delta}e^{i\gamma}\right)~,
\eea
where $\xNP \equiv A_{\rm NP}/A'_0, \tl\equiv \lambda/(1-\lambda^2/2)$. 
We are using the ``c-convention", in which the top quark in $b\to s(d)$ loop diagrams 
has been integrated out
and the unitarity relations $V^*_{tb}V_{ts(d)}=-V^*_{cb}V_{cs(d)}-V^*_{ub}V_{us(d)}$ have been 
employed. The terms $A_1$ and $-\tl A_0$ include $V^*_{cb}V_{cs}$ and $V^*_{cb}V_{cd}$,
respectively, while the terms involving $\xi'$ and $\xi$ include $V^*_{ub}V_{us}$ and 
$V^*_{ub}V_{ud}$, respectively. The weak phases $\gamma$ and $\phiNP$ change signs 
in decay amplitudes for $\bar B_s$ and $\bar B^0$.

In the U-spin symmetry limit, without neglecting any small terms such as
annihilation amplitudes, one has~\cite{Fleischer:1999pa,Gronau:2000zy}
\beq\label{Uspin}
A_1=A_0~,~~~\xi'=\xi~,~~~\delta'=\delta~.
\eeq
The first equality is susceptible to U-spin breaking correction of order
$m_s/\Lambda_{\rm QCD}$. Corrections to the  second and third relations applying to
ratios of amplitudes are expected to be smaller than $30\%$ because certain SU(3) 
breaking factors including ratios of meson decay constants and ratios of form factors 
cancel in the factorization approximation~\cite{Fleischer:1999pa}. Our discussion below 
will be restricted to CP asymmetries alone, in which the common factors in amplitudes 
$A_0$ and $\tl A_1$ cancel. Thus, our approximation relies only on the latter two relations
in (\ref{Uspin}). As we will explain, the theoretical uncertainty in extracting NP 
amplitudes is reduced further if these amplitudes occur at order $\lambda$.

We note the large enhancement by a factor $\tl^{-2}=18.6$ of the interference 
between amplitudes with weak phases $0$ and $\gamma$ in the $\Delta S=0$ process 
relative to the corresponding interference in the $\Delta S=1$ decay.  This enhancement is
effective when using {\em CP asymmetries} in the $\Delta S=0$ process for controlling the 
small amplitude $\xi'\exp(i\delta')\exp(i\gamma)$ in the $\Delta S=1$ decay.
For comparison, in earlier suggestions for using {\em decay rates}  in $\Delta S=0$ decays 
to control the small amplitude~\cite{Grossman:2003qp,Gronau:2003ep}, the effective 
enhancement factor  is only $\tl^{-1}=4.3$. This is the ratio of amplitudes with weak phase 
$\gamma$ in $\Delta S=0$ and $\Delta S=1$ decays. As we demonstrate below, the larger enhancement factor in the new method is one of two factors leading to a higher precision 
in controlling the small amplitude. A second ingredient, related to the determination 
of a strong phase difference, will be discussed below.
 
Denoting 
\bea
\ls & \equiv & e^{-i\phi_s} \frac{A(\bar B_s\to f)}{A(B_s \to f)}=
\eta_{\rm CP}e^{-i\phi_s} \frac{A(\bar B_s\to \bar f)}{A(B_s \to f)}~,\\
\lambda_{B^0} & \equiv & e^{-2i\beta}\frac{A(\bar B^0\to Uf)}{A(B^0\to Uf)}=
\eta_{\rm CP}e^{-2i\beta}\frac{A(\bar B^0\to U\bar f)}{A(B^0\to Uf)}~,
\eea
where $\eta_{\rm CP}$ is the common CP eigenvalue of $f$ and $Uf$,
it is straightforward to calculate the four CP asymmetries in the two processes,
\bea \label{CS}
C(B_s\to f) & \equiv &  \frac{1 - |\ls|^2}{1 + |\ls|^2}~,~~~~
S(B_s\to f)  \equiv \frac{2 {\rm Im}(\ls)}{1 + |\ls| ^2}~,\\
\label{CSU}
C(B^0\to Uf) & \equiv & \frac{1 - |\lambda_{B^0}|^2}{1 + |\lambda_{B^0}|^2}~,~~~~
S(B^0\to Uf)  \equiv \frac{2 {\rm Im}(\lambda_{B^0})}{1 + |\lambda_{B^0}| ^2}~.
\eea
In the U-spin symmetry limit $\xi'=\xi, \delta'=\delta$, one obtains expressions for the 
four asymmetries which are functions of  $\xi, \delta, \xNP, \delNP, \phiNP$ and the
three phases, $\beta, \gamma$ and $\phi_s$:
\bea
C(B_s\to f) & = & f_1(\gamma,\xi,\delta,\xNP, \delNP, \phiNP)~,\\
-\eta_{\rm CP}S(B_s\to f) & = & g_1(\phi_s, \gamma,\xi,\delta,\xNP, \delNP, \phiNP)~,\\
C(B^0\to Uf) & = & f_0(\gamma, \xi, \delta)~,\\
-\eta_{\rm CP}S(B^0 \to Uf) & = &  g_0(\beta, \gamma, \xi, \delta)~.
\eea

Assuming known values for the two CKM phases, $\beta, \gamma$ and the
$B_s$-$\bar B_s$ mixing phase $\phi_s$,
one is left with five parameters describing the four observables. Two of the 
parameters, $\xi$ and $\delta$, are determined from the two asymmetries 
in $B^0\to Uf$. {\em This is the proposed prescription for controlling through the latter 
process both $\xi$ and $\delta$ describing the small amplitude with weak phase 
$\gamma$ in $B_s\to f$}.
Using $\xi$ and $\delta$ as inputs in $C(B_s\to f)$ and $S(B_s\to f)$, one can 
calculate their effect on the latter asymmetries in the limit of a vanishing NP 
amplitude, $\xNP=0$. As we have pointed out, this should be more precise than 
estimating this effect by using the rate for the $\Delta S=0$ process. This follows from 
both the larger enhancement factor $\tl^{-2}$ mentioned above and information 
obtained about the strong phase $\delta$, which is unavailable when using the  rate. 

In principle, a disagreement between the predicted asymmetries in $B_s\to f$ and 
their experimental values for $\xNP=0$ would provide evidence for NP.
The discrepancy, for the predetermined values of $\xi$ and $\delta$, 
can then be used to study the three NP parameters $\xNP, \delNP$ and $\phiNP$.

We now demonstrate more explicitly the application of our proposed method,
evaluating the theoretical precision involved in identifying and potentially extracting 
the NP amplitude. For this purpose we will assume a hierarchy between the dominant  
penguin amplitude $A_1$, the smaller NP amplitude $A_{\rm NP}$ and a still smaller 
amplitude with weak phase $\gamma$,
\beq\label{hier}
1: \xNP : \xi \sim 1 : \lambda : \lambda^2~.
\eeq
We write exact expressions for $C(B^0\to Uf), S(B^0\to Uf)$ and expressions for
$C(B_s\to f), S(B_s\to f)$ which are true to order $\lambda$, keeping for illustration 
also terms of order $\xi$:
\bea
f_0 & = & \frac{2\tl^{-2}\xi\sin\delta\sin\gamma}
{1 -2\tl^{-2}\xi\cos\delta\cos\gamma + \left(\tl^{-2}\xi\right)^2}~,\\
g_0 & = & \frac
{\sin 2\beta - 2\tl^{-2}\xi\cos\delta\sin(2\beta+\gamma) + 
\left(\tl^{-2}\xi\right)^2\sin 2(\beta+\gamma)}     
{1 -2\tl^{-2}\xi\cos\delta\cos\gamma + \left(\tl^{-2}\xi\right)^2}~.\\
f_1 & = & -2\xNP\sin\phiNP\sin\delNP - 2\xi\sin\gamma\sin\delta
+ {\cal O}(\lambda^2)~,\\
g_1 & = & -\sin\phi_s + 2\cos\phi_s(\xNP\sin\phiNP\cos\delNP + 
\xi\sin\gamma\cos\delta) + {\cal O}(\lambda^2)~.
\eea 

Assuming arbitrary strong phases, one notes two interesting and useful features:
\begin{itemize}
\item The asymmetry $C(B^0\to Uf)\equiv f_0$ and the deviation of 
$-\eta_{\rm CP}S(B^0\to Uf)\equiv g_0$
from $\sin 2\beta$ are formally of order one. This is encouraging for a determination 
of $\xi$ and $\delta$ from these two asymmetries.
\item At order $\lambda$, $C(B_s\to f)\equiv f_1$ and $-\eta_{\rm CP}S(B_s\to f)\equiv g_1$ 
depend on the combination $\xNP\sin\phiNP$ and not on $\xNP$ and $\phiNP$ 
independently. This feature holds only to leading order in $\lambda$. 
At this order, $\xNP\sin\phiNP$ and $\delNP$ can be determined from 
these asymmetries when $\xi$ and $\delta$ are used as inputs obtained from 
$B^0\to Uf$.
\end{itemize}
Under our assumption (\ref{hier}),  the asymmetries $C(B_s\to f)$ and 
$-\eta_{\rm CP}S(B_s\to f)+\sin\phi_s$ are dominated by NP contributions,
$-2\xNP\sin\phiNP\sin\delNP$ and~~$2\cos\phi_s\xNP\times$
$\sin\phiNP\cos\delNP$, which are of order $\lambda$. Standard Model terms, 
$-2\sin\gamma\,\xi\sin\delta$ and $2\cos\phi_s\sin\gamma\,\xi\cos\delta$, are of order $\lambda^2$.
Consequently, theoretical errors from SU(3) breaking in the ratios of amplitudes, $\xi\sin\delta$ 
and  $\xi\cos\delta$, are diluted by another  factor $\lambda$ in the 
determination of NP quantities, $\xNP\sin\phiNP, \cos\delNP$ and $\sin\delNP$. 
Therefore, initial U-spin breaking effects of order $m_s/\Lambda_{\rm QCD}$ 
lead to very small uncertainties of order $\lambda^3$ in the extracted amplitudes.

\bigskip
\centerline{\bf III. A FEW EXAMPLES}
\bigskip

\noindent
{\bf a. Decays involving one or two pseudoscalar mesons}

\medskip
We list a few examples of pairs of U-spin related $\Delta S=1$ and $\Delta S=0$ decays to 
CP-eigenstates involving two pseudoscalars\cite{Gronau:2000zy}:
\bea
B_s\to K^+K^-~ & , & ~~~~B^0\to\pi^+\pi^-~,\\
\label{KK}
B_s\to K^0\bar K^0~~ & , & ~~~~B^0\to\bar K^0 K^0~,\\
\label{Kpi}
B^0\to K_S\pi^0~~ & , & ~~~~B_s\to K_S\pi^0~.
\eea
Time-dependent asymmetries $C$ and $S$ in $B^0\to\pi^+\pi^-$ have been measured 
at $e^+e^-$ $B$ factories~\cite{HFAG}. Similar measurements for $B_s\to K^+K^-$ are 
being planned at Fermilab and CERN. This pair of processes, usually considered 
within the Standard Model for a determination of the weak phase 
$\gamma$~\cite{Fleischer:1999pa}, can in principle also be studied 
in a broader context for potential NP effects as described above.  In  $B_s\to K^+K^-$
one has $\xi \simeq 0.2 \simeq \lambda$~\cite{Fleischer:1999pa,Gronau:2002bh} because 
this decay involves an ordinary tree amplitude. While certain SU(3) breaking corrections 
cancel in the ratio $\xi$, the hierarchy (\ref{hier}) does not hold in $B_s\to K^+K^-$. 
Therefore, SU(3) breaking corrections in the determination of a potential NP amplitude 
are not further suppressed by $\lambda$. 

Time-dependent asymmetry measurements are very challenging for the processes 
(\ref{KK}) and (\ref{Kpi}) involving only neutral pions and kaons, and will not be discussed
much further. Such measurements have been made for $B^0\to K_S\pi^0$ at 
$B$ factories~\cite{HFAG}, interpreted in terms of a CKM amplitude with weak phase 
$\gamma$~\cite{Gronau:2003kx} and in terms of potential NP 
contributions~\cite{Buras:2003dj}.  Measuring time-dependence in $B_s\to K_S\pi^0$ seems less  
feasible at hadronic colliders. Somewhat easier are decays involving corresponding 
pairs of pseudoscalar and vector mesons, e.g.,
\beq
\label{Krho}
B^0\to K_S\rho^0~~  , ~~~~B_s\to K_S\rho^0~.
\eeq
While the $\rho^0\to\pi^+\pi^-$ decay vertex permits a time-dependent measurement,
one would have to fight against a background from incidental pairs of charged pions
lying under the wide $\rho^0$. 
One may expect a cleaner signal for the pair of processes involving an $\omega$ 
instead of a $\rho^0$,
\beq
\label{Komega}
B^0\to K_S\omega~~  , ~~~~B_s\to K_S\omega~.
\eeq
  
\bigskip\noindent
{\bf b. Decays involving two vector mesons}

\medskip
Decays into two charmless neutral vector mesons, each decaying to a pair of charged
particles, are slightly more challenging than decays involving pseudoscalar mesons,
but are very interesting theoretically and experimentally. Identifying CP-eigenstates 
requires studying both the time and angular dependence for the four final decay particles.
These processes provide a high potential for probing NP effects.  
The number of amplitudes increases by a factor three relative to decays to two 
pseudoscalars due to three independent  polarization states.
The number of asymmetry observables is six times larger (see discussion below).  
The additional information permits controlling more accurately small Standard Model 
terms  and studying potential NP amplitudes with fewer ambiguities.

Consider a generic pair of $\Delta S=1$ and $\Delta S=0$ processes,
$B_s\to V_1V_2, ~B^0\to U(V_1V_2)$. The decay amplitudes for a given
polarization can be written in
the U-spin symmetry limit in analogy with (\ref{AmpBs}) and (\ref{AmpB0}),
\bea\label{AmpBs-tr}
A_k(B_s\to V_1V_2) & = & (A_1)_k\left(1 + \xi_k e^{i\delta_k}e^{i\gamma} + 
\xNPk e^{i\delNPk}e^{i\phiNP}\right)~,\\
\label{AmpB0-tr}
A_k(B^0\to U(V_1V_2)) & = & -\tl (A_0)_k\left(1 - \tl^{-2}\xi_k e^{i\delta_k}e^{i\gamma}\right)~.
\eea
The amplitudes $A_{k}$ with $(\eta_{\rm CP})_k=+1,+1,-1$, for $k=L, ||, \perp$,  
correspond to two vector mesons which are either 
longitudinally polarized ($L$), or  transversely polarized with linear polarization 
parallel ($||$) or perpendicular ($\perp$) to one another~\cite{Rosner:1990xx}.
We are assuming a single NP weak phase $\phiNP$ which is independent of 
the vector meson polarization.

Time-dependent decay distributions depend on transversity angles
defining directions for the final outgoing particles~\cite{Rosner:1990xx}.
CP asymmetries $C_{kl}$ and $S_{kl}$ ($k,l=L, ||, \perp$) multiplying $\cos\Delta 
mt$ and $\sin\Delta mt$ can be defined for each of six independent functions 
of tranversity angles~\cite{Sinha:1997zu}. The measurable asymmetries 
$C_{kl}$ and $S_{kl}$ are given by expressions analogous to 
(\ref{CS}) and (\ref{CSU}) (details will be given elsewhere~\cite{FG-upcoming}),
\bea
C_{kl} & = & \frac{{\cal F}(1-\lambda^*_k\lambda_l)}
{{\cal F}(1+\lambda^*_k\lambda_l)}~,\\
S_{kl}  & = & \frac{{\cal F}\left[i(\lambda^*_k-\lambda_l)\right]}
{{\cal F}(1+\lambda^*_k\lambda_l)}~,~~~~~~k,l=L, ||, \perp~,
\eea
where 
\bea
{\cal F} & = & \left\{ \begin{array}{c}{\rm Re}:~~k=L,~~~~~l=|| \cr
~~{\rm Im}:~~k=L,||,~~l=\perp~,
\end{array} \right.\\
\lambda_k &= & \left\{ \begin{array}{c}e^{-i\phi_s}
A_k(\bar B_s\to V_1V_2)/A_k(B_s\to V_1V_2):~~~~~~~~~B_s\\
e^{-2i\beta}A_k(\bar B^0\to U(V_1V_2))/A_k(B^0\to U(V_1V_2)):~~B^0~.
\end{array} \right.
\eea

The twelve observables $C_{kk},~S_{kk}~(k=L, ||, \perp$), combining $B_s$ and $B^0$ 
decays, are analogous to the four asymmetries  $C$ and $S$ in $\Delta S=1$ and 
$\Delta S=0$ decays to two pseudoscalars. 
They permit mutually independent  determinations for the three sets of  four 
hadronic parameters, $\xi_k, \delta_k$ and $\xNPk\sin\phiNP,\delNPk$, assuming a 
hierarchy as in (\ref{hier}).
The other twelve asymmetries, $C_{kl},~S_{kl}~(k\ne l)$, depend on the same twelve 
parameters and on four relative strong phases, two among $(A_1)_k$ and two among 
$(A_0)_k$.  This information may improve precision and resolve  discrete 
ambiguities obtained when using only $C_{kk}$ and $S_{kk}$.
Furthermore, while $C_{kk}$ and $S_{kk}$ have been shown to depend on $\xNPk\sin\phiNP$,
 ``mixed" asymmetries such as $C_{\perp i}~(i=L,||)$ depend also on 
 $\xNPk\cos\phiNP$~\cite{FG-upcoming}.  This permits determining separately the 
 magnitude $\xNPk$ and weak phase $\phiNP$ of the NP amplitude.

Examples of pairs of SU(3)-related decays to which our proposed method may be 
applied include the following:
\bea\label{K*0K*0}
B_s\to K^{*0}\bar K^{*0}~(K^{*0}\to K^+\pi^-)~~ & , & ~~~~B^0\to\bar K^{*0} K^{*0}~
(K^{*0}\to K^+\pi^-)~,\\
\label{K*0rho0}
B^0\to K^{*0}\rho^0~(K^{*0}\to K_S\pi^0)~~ & , & ~~~~B_s\to K^{*0}\rho^0~
(K^{*0}\to K_S\pi^0)~,\\
\label{phiphi}
B_s\to\phi\phi~(\phi\to K^+K^-)~~ & , & ~~~~B_s\to \phi \bar K^{*0}~
(\bar K^{*0}\to K_S\pi^0)~.
\eea

Decay rates and CP asymmetries in the first pair of  processes (\ref{K*0K*0})
have been suggested very recently as tests for consistency within the Standard 
Model~\cite{Ciuchini:2007hx}. Our method for studying potential NP amplitudes
avoids the use of decay rates, which 
are expected to introduce sizable SU(3) breaking corrections.
While $B^0\to K^{*0}\bar K^{*0}$ has already 
been observed with a branching ratio of about $0.5\times 10^{-6}$~\cite{BabarK*K*},
the decay $B_s\to K^{*0}\bar K^{*0}$ is expected to have an order of magnitude 
larger branching ratio~\cite{Beneke:2006hg,Ali:2007ff}. The contribution of an amplitude 
with weak phase $\gamma$ in this process is expected to be very small, 
$\xi\sim \lambda^2$. Thus, our suggested analysis of a potential NP amplitude 
in $B_s\to K^{*0}\bar K^{*0}$ should work well in this case.

A very interesting $B_s$ decay mode involving $\xi\sim \lambda^2$ is 
$B_s\to\phi\phi$. A handful of events have been observed in this mode a few 
years ago by the CDF collaboration at the Fermilab Tevatron, corresponding to a 
branching ratio of  $(14^{+6}_{-5}\pm 6)\times 10^{-6}$~\cite{Acosta:2005eu}. 
It is estimated that by now the signal  has grown to about 150 events~\cite{Rescigno}. 
A  proposal for studying time and angular dependence in this decay mode 
has been made by the LHCb collaboration at the CERN Large Hadron 
Collider~\cite{LHCb}. The proposal is based on an estimated sample of about 3100 
events collected in one year of running. 

While the decay mode $B_s\to\phi\phi$ has no U-spin counterpart, it may be related 
through flavour SU(3) to $B_s\to \phi \bar K^{*0}$.
The two processes involve each a penguin 
amplitude, a singlet penguin amplitude and an electroweak penguin amplitude, related to 
each other through an operator replacement $(\bar sb) \leftrightarrow (\bar db)$
and a similar quark replacement $s\leftrightarrow d$ in the final state~\cite{Gronau:1996ga}. 
An assumption about a negligible penguin 
annihilation contribution in $B_s\to\phi\phi$ may be verified by obtaining a stringent upper 
bound on $\b(B^0\to\phi\phi)$ which is dominated by penguin annihilation.
As we have noted, an SU(3) breaking correction cancels largely in the ratios $\xi_k$, and
the effect of this correction on extracting a potential NP amplitude of order 
$\lambda$ is further suppressed by $\lambda$ and can thus be neglected.
Thus, expressions analogous to (\ref{AmpBs-tr}) and (\ref{AmpB0-tr}) apply to this case 
involving $B_s\to \phi\phi$ and $B_s\to \phi \bar K^{*0}$. The first amplitude 
must be multiplied by a factor $\sqrt{2}$ to account for identical particles.

The decay $B_s\to \phi\bar K^{*0}$ is expected to have a branching ratio
of about $0.5\times 10^{-6}$~\cite{Beneke:2006hg,Ali:2007ff}. 
This corresponds to observing an order of one hundred signal events in one year at 
the LHCb. An analysis involving both time and angular dependence requires 
several years of running at the LHC.
The direct asymmetries $C_{kl}$ in $B_s\to \phi \bar K^{*0}$ can be measured through 
the self-tagging flavour state, $\bar K^{*0}\to K^-\pi^+$. The mixing-induced asymmetries $S_{kl}$ 
would have to be measured in decays to a CP eigenstate, $\bar K^{*0}\to K_S\pi^0$.
This may be challenging for experiments at hadron colliders and can also be done 
at a Super-$B$ $e^+e^-$ collider running at the 
$\Upsilon(5S)$~\cite{Baracchini:2007ei}.  The case of $B_s\to\phi\phi$ versus 
$B_s\to \phi \bar K^{*0}$ will be studied in detail elsewhere~\cite{FG-upcoming}.

\bigskip
\centerline{\bf IV. CONCLUSION}
\bigskip

We have suggested a method for studying NP amplitudes in penguin-dominated 
strangeness-changing $B_s$ decays by comparing time-dependent CP asymmetries
in these decays  to asymmetries in SU(3) related strangeness-conserving decays. 
Assuming that a NP amplitude of order $\lambda$ occurs in $\Delta S=1$ processes 
but not in $\Delta S=0$ decays, we have shown that these asymmetries determine 
with high precision small Standard Model amplitudes with weak phase $\gamma$ 
and potential NP amplitudes. Corrections from SU(3) breaking, usually assumed to 
be of order $m_s/\Lambda_{\rm QCD}$, are suppressed by two factors: 
\begin{enumerate}
\item The method depends on ratios of hadronic amplitudes in which certain 
SU(3) breaking factors cancel. 
\item The uncertainty in the extracted NP amplitude from SU(3) breaking  is 
suppressed by another factor of $\lambda$.
\end{enumerate}

 Decays into two vector mesons are particularly appealing. They 
 permit both  time-dependent and angular-dependent analyses, and can be  used 
 to extract  both the magnitudes of the NP amplitudes and 
 their strong and weak phases.  Two pairs of processes which are first on our list are 
 $B_s\to\phi\phi,~B_s\to\phi \bar K^{*0}$ and $B_s\to K^{*0}\bar K^{*0},~B^0\to \bar K^{*0}K^{*0}$.
These $\Delta S=1$ penguin-dominated $B_s$ decays  hold great 
promise for carrying out the proposed study  because of their rich polarization structure. 
Unlike $B^0\to \rho^+\rho^-$ which involves a dominant
longitudinally polarized amplitude and much smaller transverse amplitudes, 
in these $\Delta S=1$ processes the three polarization 
amplitudes are expected to have comparable magnitudes, 
 $|A_L|\sim \s|A_{||}|\sim \s|A_{\perp}|$~\cite{Beneke:2006hg,Ali:2007ff}, similar to the 
 situation observed in $B^0\to \phi K^{*0}$~\cite{HFAG}. 

\bigskip

\centerline{\bf ACKNOWLEDGMENTS}
\bigskip

We thank Ahmed Ali, Martin Beneke, Roger Forty, Franz Muheim, Tatsuya Nakada, 
Marco Rescigno and Jonathan Rosner for  very useful discussions.  
This work was supported in part by the Israel Science Foundation
under Grant No.\ 1052/04, and by the German-Israeli Foundation under
Grant No.\ I-781-55.14/2003.

\end{document}